\documentclass[prd,twocolumn,showpacs,floatfix,amsmath,nofootinbib,amssymb,floatfix]{revtex4}

\usepackage{epsfig}
\usepackage{longtable}
\usepackage{amsmath}
\usepackage{mathrsfs}
\usepackage{amssymb}
\usepackage{graphicx,color,dcolumn,booktabs,bm}
\usepackage{longtable,lscape,booktabs}
\usepackage{multirow,booktabs}
\usepackage{txfonts}
\usepackage{overpic}
\usepackage{siunitx}
\usepackage{indentfirst}
\usepackage{feynmf}   %{feynmp}
\usepackage{slashed}  %for Feynman symbols
\usepackage{cases}
\usepackage{color}
\usepackage{multirow}
\usepackage{epstopdf}
\usepackage{xcolor}
\usepackage{soul}
\usepackage{graphicx,color,dcolumn,booktabs,bm}
\usepackage{tabularx}
\usepackage[colorlinks,
            citecolor=blue,
            anchorcolor=red,
            menucolor=red,
            linkcolor=red,
            filecolor=red,
            runcolor=red,
            urlcolor=blue,
            frenchlinks=false]{hyperref}

\usepackage{ulem}
\newcommand{\blue}[1]{{\color{blue} {#1}}}

\usepackage{soul}

\begin{document}

%\begin{flushleft}
%ADP-04-17/T599 \\
%\end{flushleft}

\title{ $M$1 radiative and spin-nonflip $\pi\pi$ transitions of $B_c$ states in the Cornell potential model} 
\author{Zhi-bin Gao $^{1,2}$}
\author{Yan-yue Fan$^{1,2}$}
\author{Hao Chen$^{1,3,2}$\footnote{Contact author: chenhao\_qhnu@outlook.com}}
\author{Cheng-Qun Pang$^{1,3,2}$\footnote{Contact author: xuehua45@163.com}}
\affiliation{$^1$College of Physics and Electronic Information Engineering, Qinghai Normal University, Xining 810000, China\\$^2$Joint Research Center for Physics,
Lanzhou University and Qinghai Normal University,
Lanzhou 730000, China \\$^3$Lanzhou Center for Theoretical Physics, Key Laboratory of Theoretical Physics of Gansu Province, Lanzhou University, Lanzhou, Gansu 730000, China,
Lanzhou 810000, China}

\date{\today}

\begin{abstract}
In this paper, we mainly predict the rates of $M$1 radiative and spin-nonflip $\pi\pi$ transitions of the $B_{c}$-meson under the nonrelativistic Cornell potential model with a screening potential effect. 
We employ the numerical wave function to determine the $M$1 radiative transition widths of $B_c$ excited states and utilize the Kuang-Yan proposed method for the spin-nonflip $\pi\pi$ transitions among $B_c$ states. Our theoretical results are valuable for studying the $M$1 radiative and spin-nonflip $\pi\pi$ transition processes of $B_c$ states in experiments.

\end{abstract}
%\pacs{12.39.-x, 13.20.-v, 13.25.-k, 13.40.-f}
\maketitle
\section{Introduction}
The study of hadron spectroscopy has always been an important way for us to understand the nonperturbative properties of strong interactions, and the $B_{c}$-meson family plays a crucial role in our understanding of the strong interactions  of quantum chromodynamics (QCD). As the only conventional heavy quark mesons with distinct flavors, $B_{c}$ states provide a unique window of research significance into heavy quark dynamics. 
Comparing with the charmonium ($c\bar{c}$) and bottomonium ($b\bar{b}$) systems, the $B_{c}$-meson family is unique because of its enhanced stability brought about by the presence of two distinct heavy quark flavors, which reduces its width and prevents it from annihilating into gluons. 
Additionally, the low-lying excited $B_{c}$ states below the $BD$ ($BD^{*}$ or $B^{*}D$) can only reach the ground state through radiative decay and hadronic transition, followed by some weak decays. Therefore, the hadronic transition and radiative decay rates include almost the total decay width of the lowest excited $B_{c}$ states. However, experimental data on $B_{c}$ states are still scarce and require more observation and exploration to discover and understand their properties.

Over the past few decades, {\color{black}there has been some progress in experimental research on the $B_{c}$-meson family, but the processes have not been smooth sailing.} In 1981, it was predicted that the $B_{c}$ state would be composed of a quark-antiquark pair with bottom-charm \cite{Eichten:1980mw}. Afterwards, it was proposed in Refs.~\cite{Chang:1992bb,Chang:1992jb} that $B_{c}$-mesons can be detected through hadron collider experiments. A few years later, the CDF Collaboration observed a $B_{c}$-meson with the mass of $ M=(6.40\pm 0.39\pm 0.13)$ GeV by the Tevatron collider in 1998 \cite{CDF:1998ihx,CDF:1998axz},
which attracted people to carry out experimental research on the $B_{c}$-meson family. 

However, no further significant discoveries were made until 2014, 
when the ATLAS Collaboration  identified a peak at $6842\pm4\pm5 $ MeV \cite{ATLAS:2014lga},
which could be interpreted as a $B_{c}^{*}(2^3S_{1})$ excited state or a pair of analytic peaks resulting from decays of $B_{c}(2^{1}S_{0}) \rightarrow B_{c}(1^{1}S_{0})\pi^{+}\pi^{-} $ 
and $B_{c}^{*}(2^{3}S_{1}) \rightarrow B_{c}^{*}(1^{3}S_{1})\pi^{+}\pi^{-} $ followed by  $B_{c}(1^{3}S_{1}) \rightarrow B_{c}(1^{1}S_{0})\gamma $. Nevertheless, this information was not confirmed by the LHCb Collaboration with its 8 TeV data sample until 2018 \cite{LHCb:2017rqe}. {\color{black}Even so,} an experimental {\color{black}value} is provided for the transition process of $B_{c}(2^{1}S_{0}) \rightarrow B_{c}(1^{1}S_{0})\pi^{+}\pi^{-} $ in PDG \cite{ParticleDataGroup:2022pth}.
% Up to 2019, the CMS \cite{CMS:2019uhm} and LHCb \cite{LHCb:2019bem} Collaborations observed consistent signals emitted by $B_{c}(2S)$ and $B_{c}^{*}(2S)$ states in the $B_{c}(1S)\pi^{+}\pi^{-}$ invariant mass spectrum.
In the $B_{c}(1S)\pi^{+}\pi^{-}$ invariant mass spectrum, the CMS \cite{CMS:2019uhm} and the LHCb \cite{LHCb:2019bem} Collaborations observed consistent signals emitted by $B_{c}(2S)$ and $B_{c}^{*}(2S)$ states till 2019.   
It is anticipated that the upgrade of the Large Hadron Collider (LHC) will provide more data on $B_{c}$-mesons in the future, allowing a complete $B_{c}$-meson family to be constructed. {\color{black}Because of limited experimental data on $B_{c}$ states, there is little theoretical research on the decay processes of  $B_{c}$ states such as $B_{c}(2^{1}S_{0}) \rightarrow B_{c}(1^{1}S_{0})\pi^{+}\pi^{-} $ and $B_{c}^{*}(2^{3}S_{1}) \rightarrow B_{c}^{*}(1^{3}S_{1})\pi^{+}\pi^{-} $. Further theoretical research  for more $B_{c}$-mesons is necessary.}  

Recently, the BESIII Collaboration provided that the upper limit on the product branching fraction ${\cal B} (\psi(2S) \to \gamma\eta_{c}(2S))\times{\cal B}(\eta_{c}(2S)\to \pi^{+}\pi^{-}\eta_{c})$ is determined to be $2.21\times 10^{-5}$ at the 90$\%$ confidence level, which is a significant result in searching for the decay process of $\eta_{c}(2S)\to \pi^{+}\pi^{-}\eta_{c}$ ~\cite{BESIII:2024mdm}. Among them, the ${\cal B} (\psi(2S) \to \gamma\eta_{c}(2S))$ process has an important contribution. 
%It suggests that while investigating the two pion hadronic transition, the influence of M1 radiation decay cannot be disregarded. 
Thus, {\color{black}the product branching fraction ${\cal B} (B_{c}(2^3S_{1}) \to \gamma B_{c}(2^1S_{0}))\times{\cal B}(B_{c}(2^1S_{0})\to \pi\pi B_{c}(1^1S_{0}))$} may also be crucial in unraveling the mysteries of the two pion hadronic transition experiment to explore the $B_{c}$ states.
% despite the lack of information available about it. 
%This is worth exploring it for us.

{\color{black}On the theoretical side, extensive studies on low-lying states of $B_{c}$-mesons have been carried out in the past few decades
\cite{Godfrey:1985xj,Eichten:1994gt,Zeng:1994vj,Gershtein:1994dxw,Ebert:2002pp,Godfrey:2004ya,Fulcher:1998ka,AbdEl-Hady:2005wtn,Devlani:2014nda,Monteiro:2016ijw,Soni:2017wvy,Monteiro:2016rzi,Eichten:2019gig,Baldicchi:2000cf,Tang:2018myz,Ikhdair:2003tt,Ikhdair:2003ry,Li:2004gu,Wei:2010zza,Guo:2008he,Badalian:2009cx,Wang:2012kw,Chen:2011qu,Rai:2008sc,Patel:2008na,Bernotas:2008bu,Ikhdair:2004hg,AbdEl-Hady:1998uiq,Motyka:1997di}. Specifically, the low-lying vector $B_{c}$-meson was studied in the $B^{*}_{c} \rightarrow B_{c}+\gamma$, $B^{*}_{c}\rightarrow \ell + \nu_{\ell}$ and $B^{*}_{c} \rightarrow J/\psi + nh$ processes within effective theory by the helicity decomposition method which is a very important research in this area \cite{Geng:2023ffc}. In recent years, there has also been some progress in the study of highly excited $B_{c}$ states \cite{Li:2019tbn,Li:2022bre,Li:2023wgq}. At present, only a small amount of research in the study of the two pion hadronic transition of $B_{c}$ states has occurred \cite{Godfrey:2004ya,Martin-Gonzalez:2022qwd,Li:2023wgq}. The nonrelativistic quark model plays an important role in predicting the energy spectra of low-lying and highly excited states of the $B_{c}$ states \cite{Monteiro:2016ijw,Li:2019tbn,Li:2022bre}. Therefore, it is a good choice for us to use this model to further study the two pion hadronic transition of $B_{c}$ states.}

%In the case of hadronic transitions, the release of light hadrons from a high energy level transition to a low energy level is referred to as the hadronic transition process.
%Since the energy difference between the initial and the final $B_{c}$ states is expected to be minimal, which leads the emitted light hadron energy is insufficient to calculate the hadronic transition process with precision. 
%In terms of calculating the low-energy hadron transitions of heavy quarks, since 1977, Gottfried pointed out the QCD multipole expansion method \cite{Gottfried:1977gp} to solve that the emitted light hadron energy is insufficient to calculate the hadronic transition process with precision. 
 We calculate the hadronic transition by the means of the QCD multipole moment expansion
that has been studied by many scholars \cite{Gottfried:1977gp,Bhanot:1979af,Peskin:1979va,Bhanot:1979vb,Voloshin:1978hc,Voloshin:1980zf}, which has been validated. In 1981, Kuang and Yan collaborated to put forward a reasonable approach of the intermediate state and provided a practicable method for calculating the hadronic transition for the first time \cite{Kuang:1981se}. In the subsequent research, a series of studies on hadronic transitions of heavy quark systems are carried out using the Kuang-Yan approach \cite{Kuang:1988bz,Kuang:1989ub,Kuang:2002hz,Kuang:2006me,Segovia:2016xqb,Segovia:2015raa,Segovia:2014mca,Wang:2018rjg,Godfrey:2004ya,Martin-Gonzalez:2022qwd,Li:2023wgq}. Therefore, the Kuang-Yan approach is successful in calculating the spin-nonflip $\pi\pi$ transition of low excited states, which lays the foundation for us to utilize this model. At present, there is still very little experimental data on the hadronic transition of $B_{c}$ states, and it is worthwhile for us to study $\pi\pi$ transitions of $B_{c}$ states.

In this paper, we extend our study of the previous $B_{c}$ mass spectrum \cite{Li:2022bre} to include both the spin-nonflip $\pi\pi$ transitions and $M$1  radiative transitions. The theoretical framework is a nonrelativistic Cornell potential model with a screening potential effect. In addition, the $B_{c}$ states that are discussed are all below or near the $BD$ threshold \cite{ParticleDataGroup:2020ssz}.

This paper is organized as follows. In Sec.~\ref{sec:mx}, the potential model used to obtain the mass spectrum and accurate wave function of the $B_{c}$ states is introduced; {\color{black}additionally}, theoretical methods for the $M$1 radiative and hadronic transitions were provided. In Sec.~\ref{sec:results}, the analysis of calculation results of the spin-nonflip $\pi\pi$ and the $M$1  radiative {\color{black}transitions is performed, and} the reliability of our data is measured by comparing the data. In Sec.~\ref{sec:summary}, we provide a summary and give our results.

\section{Theoretical models}
\label{sec:mx}

{\color{black}In this section, we will introduce the models we used, the potential model, the magnetic dipole radiative transition, and the method of the spin-nonflip $\pi\pi$.}
\subsection {Potential model}
\label{sec:spect}

{\color{black}The mass spectrum of $B_{c}$ states is calculated by using the nonrelativistic Cornell potential model with a screening potential in Ref. \cite{Li:2022bre}. The mass spectra of intermediate hybrid states also depend on the Cornell potential model. In addition, our calculations consider the spin mixing of natural  states of $B_{c}$ states.}
%, we have also provided a brief introduction to the method of calculating mixing angles.}
Here, we briefly introduce the process of calculating the mixing angles for the $B_{c}$ states. For specific potential model processes, please refer to Appendix~\ref{sec:TmoMS}.

{\color{black}The Hamiltonian of the $B_{c}$-meson, the $L-S$ coupling term includes symmetric and antisymmetric parts, where the antisymmetric part leads to spin mixing of $B_{c}$ states. 
%The symmetric part is zero according to the calculation, 
The Hamiltonian of the antisymmetric part is expressed as}
\begin{align}
H_{\mathrm{anti}}=\frac{1}{4}\left[\left(\frac{4}{3} \frac{\alpha_{s}}{r^{3}}-\frac{b e^{-\mu r}}{r}\right)\left(\frac{1}{m_{1}^{2}}-\frac{1}{m_{2}^{2}}\right) {(\vec{S}_1-\vec{S}_2)\cdot{\vec{L}}}\right]. 
\end{align}
The mixture of states denotes
\begin{align}
L^{\prime}={ }^{1} L_{J} \cos \theta+{ }^{3} L_{J} \sin \theta, \\
L=-{ }^{1} L_{J} \sin \theta+{ }^{3} L_{J} \cos \theta,
\end{align}
where $\theta$ is the mixing angle.

\subsection{$M$1 radiative transition}

Radiative transitions in heavy quarkonium play a vital role as they not only serve as the primary decay channels for particles below the open-flavor threshold,
but also aid in a better understanding of a quarkonium's internal structure, 
including wave functions and $Q\bar{Q}$ interactions \cite{Wang:2018rjg}.
The $M$1 radiative transition's partial widths with spin-flip can be expressed as \cite{Novikov:1977dq}\ 
(from the initial state $i$ to the final state $f$)

\begin{equation}
\Gamma( i\to f \gamma) = \frac{\alpha}{3} \;\delta_{S S^{\prime}\pm1} \mu^2 \; \omega^3 \frac{2J_f +1}{2L+1}
 \, |\,\langle f | \; j_0(\omega r/2) \; |\,  i \rangle\, |^2,  
\end{equation}
where 
\begin{equation}
\mu= \frac{e_c}{m_c} - \frac{e_{\bar{b}}}{m_{\bar{b}}},
\end{equation}
and
\begin{equation}
    \langle f |  j_0(\omega r/2)  | i \rangle=\int_{0}^{\infty}R_{n^{\prime}L^{\prime}}(r)j_{0}(\omega r/2)R_{nL}(r)r^2dr.
\end{equation}
Parameters $e_c$ and $e_{\bar{b}}$ denote the charges of the $c$-quark and $\bar{b}$-antiquark, respectively, in units of $|e|$.
Specifically, $e_c$ is equal to $2/3$ and $e_{\bar{b}}$ is equal to $-1/3$. 
Furthermore, $m_c$ and $m_{\bar{b}}$ refer to the masses of the quarks that were previously mentioned. $\alpha$ is the fine structure constant, which is a dimensionless parameter that typically takes $\alpha\approx\frac {1}{137}$.
The term $j_{0}(\omega/2)$ is the spherical Bessel function and $\omega$ is the energy of the photon. 

{\color{black}According to the conservation of energy and momentum, the energy of photon can be obtained from}
\begin{align}
\label{eq:gze}
    M_{i}=\sqrt{M_{f}^{2}+\omega^{2}}+\omega,
\end{align}
where $M_{i}$ and $M_{f}$ are the masses of the initial and final states of the $B_{c}$ states, respectively.

\subsection{Hadronic transition}
\label{subsec:HTT}
%\sout{The hadronic transition is an important decay method for studying $B_{c}$ states,
%which can help us understand more about the properties of $B_{c}$ states.}
For $B_{c}$ states, the hadronic transition of $B_{c}$ states is a process in which a light hadron is released when the $c\bar{b}$ state transitions to a lower energy level.
It can be given by 
\begin{equation}
\Phi_{i} \to \Phi_{f}+h,
\end{equation}
where $\Phi_{i}$ and $\Phi_{f}$ are defined as the initial and final states of $B_c$ states, respectively, and $h$ denotes the emitted light hadron(s) which are kinematically dominated by either {\color{black}single meson ($\pi^{0}$, $\eta$, $\omega$, $\ldots$) or two mesons ($2\pi$, $2K$, $\ldots$).} 

Since the difference in mass between the initial and the final states is small, the momentum of the light hadron(s) $h$ is correspondingly low.
Without taking into account the coupling channel effect, the light hadron(s) $h$ is converted from gluons emitted by the quark or antiquark, so the momentum of the emitting gluons is low as well. Therefore, this process cannot be calculated using perturbation QCD. Gottfried pointed out in Ref.~\cite{Gottfried:1977gp} that this situation can be solved by the method of multipole expansion since the wavelengths of emitted gluons are larger than the size of $B_c$-meson states. After the expansion of the gluon field, the Hamiltonian of the system can be given by \cite{Kuang:2006me}
\begin{equation}
{\cal H}^{\rm eff}_{\rm QCD} = {\cal H}^{(0)}_{\rm QCD} + {\cal H}^{(1)}_{\rm
QCD},
\label{eq:Hqcd}
\end{equation}
with ${\cal H}^{(0)}_{\rm QCD}$ the sum of the kinetic and potential energies of the bottom-charmed meson, and ${\cal H}^{(1)}_{\rm QCD}$ are defined as 
\begin{equation}
\begin{split}
{\cal H}^{(1)}_{\rm QCD} &={\cal H}^{(1)}+{\cal H}^{(2)},\\
{\cal H}^{(1)} &= Q_{a} A^{a}_{0}(x,t), \\
{\cal H}^{(2)} &=-d_{a} E^{a}(x,t) - m_{a} B^{a}(x,t),
\label{eq:Hqcd2}
\end{split}
\end{equation}
where $Q_{a}$ corresponds to the color charge, $d_{a}$ to the color-electric dipole moment, and $m_{a}$ to the color magnetic dipole moment. Since we are working with $c\bar{b}$ pairs that form a color singlet object, there is no contribution from the ${\cal H}^{(1)}$ and only $E_{l}$ and $B_{m}$ transitions can take place. The lowest order term between two color singlets involves two gluons, and therefore the lowest multipole is the double electric-dipole term (E1-E1).

Next, we will give the brief outline of the processes involved in the spin-nonflip $\pi\pi$ transitions. For further details on specific processes, refer to Ref. \cite{Kuang:2006me}.

%\subsubsection{Spin-nonflip \texorpdfstring{$\pi\pi$}{}}
%\label{subsubsec:Spinnonflip}

The spin-nonflip $\pi\pi$ transitions of interest in this paper are mainly E1-E1, and {\color{black}the transition amplitude is obtained from the $S$-matrix elements} given in Ref. \cite{Kuang:2006me},
\begin{equation}
{\cal M}_{E1E1}=i\frac{g_{E}^{2}}{6} \left\langle\right.\!\! \Phi_{f}h \,
|\vec{x}\cdot\vec{E} \, \frac{1}{E_{i}-H^{(0)}_{QCD}-iD_{0}} \,
\vec{x}\cdot\vec{E}| \, \Phi_{i} \!\! \left.\right\rangle,
\label{eq:E1E1}
\end{equation}
where $g_{E}$ is the coupling constant for electric dipole (E1) gluon emission, $\vec{x}$ is the separation between the quark and antiquark, $\vec{E}$ is the color-electric field, and $G(E_{i})= \frac{1}{E_{i}-H^{(0)}_{QCD}-iD_{0}}$ is Green's function, {\color{black}$(D_0)_{bc}\equiv\delta_{bc}\partial_{0}-g_{s}f_{abc}A^{a}_{0}$ in $\frac{1}{E_{i}-H^{(0)}_{QCD}-iD_{0}}$.} 

%$\left|\Phi _{i}\right\rangle$ is the initial $c\bar{b}$ state and $\langle\Phi_{f}h|$ is the final $c\bar{b}$ state and a light hadron.

After inserting a complete set of intermediate states and using a quark confining string (QCS) model, the transition amplitude in Eq.~\eqref{eq:E1E1} can be written as

\begin{equation}
{\cal M}_{E1E1}=i\frac{g_{E}^{2}}{6} \sum_{kl}
\frac{\left\langle\right.\!\! \Phi_{f}|x_k|kl \!\!\left.\right\rangle
\left\langle\right.\!\! kl|x_l|\Phi_i \!\!\left.\right\rangle}{E_I-E_{kl}}
\left\langle\right.\!\! \pi\pi|E^{a}_{k} E^{a}_{l}|0 \!\!\left.\right\rangle,
\label{eq:factorizedE1E1}
\end{equation}
where $E_{kl}$ is the energy eigenvalue of the intermediate state $|kl\rangle$ with the principal quantum number $k$ and the orbital angular momentum $l$ and corresponding eigenvalues in the sector of the lowest string excitation, $E^{a}$ is the color electric field.

The intermediate states in the hadronic transition consist of a gluon and a color-octet $c\bar{b}$ that are the states after the emission of the first gluon and before the emission of the second gluon. Thus, these states are the so-called hybrid states. A rational model is needed to solve these states, which cannot be calculated from the first principles of QCD. In fact, we shall take the QCS model that has already been used for the study of similar hadronic transitions in the charmonium and bottomonium sectors~\cite{Segovia:2014mca, Segovia:2015raa, Segovia:2016xqb}, and this will be explained later.

The transition amplitude can be divided into two parts from  Eq.~(\ref{eq:factorizedE1E1}), which are a heavy quark multipole gluon emission (MGE) factor (the summation) and an H (hadronization) factor $\left\langle\pi\pi|E^{a}_{k} E^{a}_{l}|0\right\rangle$, respectively. Using the eigenvalues and wave functions of the intermediate hybrid mesons and the initial and the final quarkonium states, the MGE factor can be calculated.
The H factor reflects the conversion of the two emitted gluons into light hadrons after hadronization. Because of its low energy, it is highly nonperturbative so that this matrix element cannot be calculated with perturbative QCD. In this case, phenomenological methods based on the technology of the soft pion should be applied\cite{Brown:1975dz}. In the center-of-mass frame, the two pion momenta $q_{1}$ and $q_{2}$ are the only independent variables describing this matrix element that can be written as\cite{Brown:1975dz, Yan:1980uh, Kuang:1981se, Kuang:2006me} 
\begin{equation}
\begin{split}
& \frac{g_{E}^{2}}{6} \left\langle\right.\!\!
\pi_{\alpha}(q_{1})\pi_{\beta}(q_{2})|E^{a}_{k}E^{a}_{l}|0
\!\!\left.\right\rangle =
\frac{\delta_{\alpha\beta}}{\sqrt{(2\omega_{1})(2\omega_ {2})}} \,\times \\
&
\times
\left[C_{1}\delta_{kl}q^{\mu}_{1}q_{2\mu} + C_{2}\left(q_{1k}q_{2l}+q_{1l}q_{2k}
-\frac{2}{3}\delta_{kl}\vec{q}_{1}\cdot\vec{q}_{2}\right)\right],
\label{eq:c1c2}
\end{split}
\end{equation}
where $C_{1}$ and $C_{2}$ are two unknown constants that are related to our ignorance about the mechanism of the conversion of the emitted gluons into light hadron(s) and  $q^{\mu}_{1}$ and $q_{2\mu}$ are momentum components. The $C_{1}$ term is isotropic, {\color{black}and} the $C_{2}$ term has a $l=2$ angular dependence. Thus, $C_{1}$ is involved in hadronic transitions where $\Delta l = l_f-l_i = 0$, while $C_{2}$ begins to participate when $\Delta l = 2$. The specific calculation processes of the spin-nonflip $\pi\pi$ transition can be found in Appendix~\ref{sec:st}.

%\subsubsection{A model for hybrid mesons}
%\label{subsubsec:hybrid}
In this article, the intermediate hybrid state is described by the QCS model \cite{Tye:1975fz,Giles:1977mp,Buchmuller:1979gy}. 
 The specific introduction and content of the effective potential is provided in Appendix~\ref{sec:ep}.

Another important characteristic of the hybrid states is that their mass spectrum has a threshold: once a certain threshold is reached, no more states can be found. Hybrid meson masses calculated in the $B_c$ sector  are shown in Table~\ref{tab:hybridsbb}.

\begin{table}[!t]
% \begin{tabularx}{\textwidth}{X|X|X}
    \begin{center}
    \caption{\label{tab:hybridsbb} Hybrid meson masses of the $c\bar{b}$ sector, in MeV.}
    \renewcommand\arraystretch{1.3}
    \begin{tabular}{c c c c}
    \toprule[1.0pt]\toprule[0.5pt]
   ~~~~~~~~ k~~~~~~~ & ~~~~~~~~$l=0$~~~~~~~~~ & ~~~~~~~~~$l=1$~~~~~~~ &~~~~~ $l=2$ ~~~~~~\\
    \hline
    $1$  & $7254$ & $7556$ & $7730$ \\
    $2$  & $7634$ & $7820$ & $7959$ \\
    $3$  & $7893$ & $8039$ & $8155$ \\
    $4$  & $8105$ & $8227$ & $8328$ \\
    $5$  & $8285$ & $8392$ & $8481$ \\
    $6$  & $8447$ & $8539$ & $8618$ \\
    $7$  & $8589$ & $8670$ & $8740$ \\
    $8$  & $8716$ & $8788$ & $8851$\\
    $9$  & $8829$ & $8893$ & $8947$\\
    $10$ & $8932$ & $8988$ & $9040$\\
    $11$ & $9023$ & $9074$ & $9116$\\
    $12$ & $9106$ & $9151$ & $9205$\\
    $13$ & $9178$ & $9215$ & $9273$\\
    $14$ & $9251$ & $9269$ & $9332$\\
    $15$ & $9313$ & $9325$ & $9359$\\
    $16$ & $9351$ & $9404$ & $9464$\\
    $17$ & $9449$ & $9494$ & $9512$\\
    $18$ & $9505$ & -      &  -     \\
    \hline
    \multicolumn{4}{c}{Threshold = 9531 MeV} \\
    \bottomrule[0.5pt]\bottomrule[1.0pt]
  \end{tabular}
    \end{center}
  %   \end{tabularx}
    \end{table}

\section{Numerical results and phenomenological analysis}
\label{sec:results}
{\color{black}In this section, we analyze the results of the spin-nonflip $\pi\pi$ transition and $M$1 radiative transition, predict the possibility of the spin-nonflip $\pi\pi$ transition observation by comparing data, and provide the value of the product of some branching fractions, e.g.,  ${\cal B} (B_{c}(2^3S_{1}) \to \gamma B_{c}(2^1S_{0}))\times{\cal B}(B_{c}(2^1S_{0})\to \pi^{+}\pi^{-}B_{c}(1^1S_{0}))$.}

\subsection{The analysis of spin-nonflip \texorpdfstring{$\pi\pi$}{} transitions}
\label{sec:AOS}
%{\color{black}We have used the QCD Multipole Expansion (QCME) method to calculate the spin non-flip $\pi\pi$ transition. This method mainly consists of two parts: one is the heavy quark multipole gluon emission process, and the other is the process {of gluon emission and light hadronization.} 
%\sout{from gluon emission to light hadronization.} 
%Among them, the hybrid state is composed of the first gluon and a color-octet state, which is calculated using the QCS model, and the specific process is shown in the Appendix~\ref{sec:ep}.}

We use the theoretical model under the framework of gauge invariant QCD multipole expansion (QCDME) and the Kuang-Yan approach introduced in Sec. \ref{subsec:HTT} to calculate the spin-nonflip $\pi\pi$ transition. The mass spectrum and the wave function of $B_{c}$ states are obtained in Ref.~\cite{Li:2022bre}, and we have calculated the mass spectrum and the wave function of the hybrid states using parameters given in Table~\ref{Parameters}. Then we only need to determine the two unknown parameters $C_{1}$ and $C_{2}$ in Eq.~\eqref{eq:c1c2}. The two parameters are described as Wilson coefficients, which depend on the characteristic energy scale of the physical process. In fact, the above two parameters $C_{1}$ and $C_{2}$ depend on the partial hadron transition experimental width. For the case of the hadron transition of $B_{c}$ states, the results obtained by taking into account both bottomonium and charmonium are maybe more reliable \cite{Mannel:1995jt,Novikov:1977cm,Godfrey:2004ya} than those only considering bottomonium\cite{Martin-Gonzalez:2022qwd}.
In the following calculation, we take both bottomonium and charmonium into account and fit their transition rates with the method used in  Ref. \cite{Godfrey:2004ya}. The amplitudes for E1-E1 transitions depend quadratically on the interquark separation so the scaling law between a $c\bar{b}$ rate and the corresponding $Q\bar{Q}$ rate is given by \cite{Yan:1980uh}
\begin{equation}
\frac{\Gamma(c\bar{b})}{\Gamma(Q\bar{Q})}=
{\frac{ \langle r^2 (c\bar{b}) \rangle^2}
{ \langle r^2 (Q\bar{Q}) \rangle^2}}.
\end{equation}

In Table \ref{tab:H-S}, we provide the scaling factors that relate the input width to the width of $c\bar{b}$. Among them, the values of input width ($Q\bar{Q}$) are calculated from the total width and transition branch ratio given in PDG \cite{ParticleDataGroup:2022pth}. Specifically, as mentioned in Ref. \cite{Godfrey:2004ya}, the input width of the process  $\Upsilon(1D) \to \Upsilon(1S)\pi\pi$ is difficult to calculate accurately due to the lack of experimental values. In our calculation, we use the transition rate of process $\Upsilon_{2}(1D) \to \Upsilon(1S)\pi^{+}\pi^{-}$ given in Ref. \cite{Brambilla:2010cs} as input, with a value of 0.188 keV. For the spin-nonflip $\pi\pi$ case, because of the lack of experimental values for the process of $\Upsilon_{2}(1D) \to \Upsilon(1S)\pi^{0}\pi^{0}$, we take half of the transition rate of process $\Upsilon_{2}(1D) \to \Upsilon(1S)\pi^{+}\pi^{-}$. And then we obtained the transition rate of process $\Upsilon(1D) \to \Upsilon(1S)\pi\pi$ as shown in Table \ref{tab:H-S}. The $C_{1}$ term is isotropic and contributes to the transition between $S$ and $S$ waves, while the $C_{2}$ term has a $l=2$ angular dependence and contributes to the transition from $D$ waves to $S$ waves. Therefore, the parameter $C_{1}$ is obtained from the process $B_{c}(2^3S_1 \to 1^3S_1 + \pi\pi)$ as input, the parameter $C_{2}$ is obtained from the process $B_{c}(1^3D_1 \to 1^3S_1 + \pi\pi)$ as input. By using the fitted input values in Table \ref{tab:H-S}, we can obtain the numerical results for $C_{1}$ and $C_{2}$, respectively. For the spin-nonflip $\pi\pi$ case
\begin{equation}
    \label{eq:c1-1}
    \begin{split}
    |C_{1}|^2&=8.656 \times 10^{-5},\\
    |C_{2}|^2&=1.712 \times 10^{-4}.
    \end{split}
\end{equation}

It should be noted that mixing is involved in our calculations, so the process of $B_{c}(1^3D_2 \to 1^3S_1 + \pi\pi)$ is not considered as an input for determining parameter $C_{2}$.
\begin{table}[htbp]\footnotesize
	\centering
\caption{The parameters in the potential model adopted in this work.}
\label{Parameters}
\[\begin{array}{ccccccc}
\toprule[1.0pt]\toprule[0.5pt]
\text{Parameter} & \text{Value} & \text{Parameter} & \text{Value}  &\\
\midrule[0.5pt]
m_b    &  \color{black}{5.368}~\text{GeV}   & m_u,\,m_d & \color{black}{0.606}~\text{GeV}  &\\
m_c    &  \color{black}{1.984}~\text{GeV}  & m_s    &  \color{black}{0.780}~\text{GeV}  & \\
\alpha_s & \color{black}{0.3930}  & \sigma & \color{black}{1.842}~\text{GeV}  &  \\
b       & \color{black}{0.2312}~\text{GeV}^2  & c   & \color{black}{-1.1711}~\text{GeV}  &  \\
\mu & \color{black}{0.0690}~\text{GeV}  &\color{black}{ r_c }&\color{black}{0.3599 ~\text{GeV}^{-1} }& \\
\bottomrule[0.5pt]\bottomrule[1.0pt]
\end{array}\]\label{tab1}
\end{table}

\begin{table*}
\caption{{The hadronic transition input width of the $B_{c}$ states fitted by charmonium and bottomonium.} 
\label{tab:H-S}}
%\begin{center}
\begin{ruledtabular}
\begin{tabular}{l l c c} %\hline
Transition & $(Q\bar{Q}):$   Rate [keV] 
	& $\langle r^2(c\bar{b})\rangle / \langle r^2(Q\bar{Q})\rangle$
	& Reduced $c\bar{b}$ rate [keV] \\
\hline 

$2^3S_1 \to 1^3S_1 + \pi\pi$ & $(c\bar{c})$: $156\pm4$ \footnotemark[1]
    		& 0.7 &$76\pm2$	\\
			& $(b\bar{b})$:  $8.46\pm0.7$ \footnotemark[1]
		& 1.86 & $29.27\pm2.42$	\\
		& Average	&	& $53\pm2$ \\

$1^3D_1 \to 1^3S_1 + \pi\pi$ & $(c\bar{c})$:  $74.3\pm3$ \footnotemark[1]
		& 0.65 & $31.4\pm1$	\\
		& $(b\bar{b})$:  $0.413$ \footnotemark[3]
		& 1.61 &$1.071$	\\
		& Average	&	& $16.2\pm0.5$ \\
$1^3D_2 \to 1^3S_1 + \pi\pi$ & $(c\bar{c})$:  $123.5$ \footnotemark[1]
		& 0.72 & $64$	\\
			& $(b\bar{b})$:  $0.284$ \footnotemark[2]
		& 1.68 & $0.531$	\\
		& Average	&	& $33$\\

\end{tabular}
\end{ruledtabular}
\footnotetext[1]{From PDG Ref.\cite{ParticleDataGroup:2022pth}.}
\footnotetext[2]{From $\Gamma(1^3D_2 \to 1^3S_1 + \pi^{+}\pi^{-})=0.188$ keV in Ref. \cite{Brambilla:2010cs}.}
\footnotetext[3]{From $\Gamma(1^3D_2 \to 1^3S_1 + \pi\pi)=0.284$ keV as an input.}

\end{table*}

\begin{table}[htbp]\centering
        \caption{Decay rates of spin-nonflip $\pi\pi$ transitions between $B_c$ states. Here, mixing angles $\theta_{1P}=-24.3^{\circ}$, $\theta_{2P}=-28.4^{\circ}$, $\theta_{1D}=-41.7^{\circ}$ \cite{Li:2022bre}, IS, FS, and TW represent  initial states, final states, and this work, respectively; the decay rates are given in units of keV.}
        \renewcommand\arraystretch{1.2}
        \label{E1-E1 transitions}
        \begin{tabular*}{84mm}{c@{\extracolsep{\fill}}lcccc}
        \toprule[1.0pt]\toprule[0.5pt]
        \text{IS} & \text{FS}  &\text{TW}& \text{MGI \cite{Li:2023wgq}} & \text{GI \cite{Godfrey:2004ya}} &\text{\cite{Martin-Gonzalez:2022qwd}} \\
        \toprule[0.5pt]
        $2^1S_0$    & $1^1S_0 + \pi\pi$ &46& 25 & 57 & 42 \\
        \hline
        $2^3S_1$    & $1^3S_1 + \pi\pi$ &53& 21 & 57 & 41 \\
        \hline
        \multirow{3}*{$2^3P_0$}    & $1^3P_0 + \pi\pi$ &104& 2.8 & 0.97 & 12 \\
                                   & $1P_1 + \pi\pi$  &0 &0 & 0 & 0 \\
                                   & $1P_1^{\prime} + \pi\pi$  & 0 &0 & 0 & 0 \\
                                   & $1^3P_2 + \pi\pi$ & 0.029 & $1.2\times10^{-4}$ & 0.055 & $5.5\times10^{-3}$ \\ \hline   
    
        \multirow{3}*{$2P_1$}     & $1^3P_0 + \pi\pi$ &0&0 & 0 & 0\\
                                   & $1P_1 +  \pi\pi$  &0.08& 1.5 & 2.7 & 11\\
                                   & $1P_1^{\prime} +  \pi\pi$  &0.020& 0.77 & 0.020 & \\
                                   & $1^3P_2 + \pi\pi$ &0.191& $6.3\times10^{-4}$ & 0.037 & 0.012\\
                                   \hline 
        \multirow{3}*{$2P_1^{\prime}$}  & $1^3P_0 + \pi\pi$ &0& 0 & 0 & 0\\
                                   & $1P_1 +  \pi\pi$  &0.015& 1.4 & 0.10 & \\
                                   & $1P_1^{\prime} +  \pi\pi$  &0.004& 1.6 & 1.2 & 11\\
                                   & $1^3P_2 + \pi\pi$ &0.034& $2.7\times10^{-4}$ &$4.0\times10^{-3}$ & \\
                                   \hline

        \multirow{3}*{$2^3P_2$}    & $1^3P_0 + \pi\pi$   &0.350& $5.7\times10^{-3}$ & 0.011 & 0.018\\
                                   & $1P_1 +  \pi\pi$  &0.215&  $2.7\times10^{-3}$& 0.021 & 0.020\\
                                   & $1P_1^{\prime} +  \pi\pi$  &0.052&  $9.7\times10^{-4}$ & $4.0\times10^{-3}$ & \\
                                   & $1^3P_2 + \pi\pi$   &8.9&3.0 & 1.0 & 11\\
                    \hline
       
                   $1^3D_1$    & $1^3S_1 + \pi\pi$ &16.2& 0.15 & 4.3 & 0.75 \\
                   \hline
        \multirow{2}*{$1D_2$}    & $1^1S_0 + \pi\pi$ &9.4& 0.20 & 2.1 &  \\
                                   & $1^3S_1 + \pi\pi$ &9.0& 0.066 & 2.2 & \\
                                   \hline
        \multirow{2}*{$1D_2^{\prime}$}    & $1^1S_0 + \pi\pi$ &13.3& 0.12 & 2.2 &  \\
                                        & $1^3S_1 + \pi\pi$ &8.2&0.13 & 2.1 &  \\
                   \hline
                   $1^3D_3$    & $1^3S_1 + \pi\pi$ &17.4& 0.23 & 4.3 & 0.84 \\
       
        \bottomrule[0.5pt]\bottomrule[1.0pt]
        \end{tabular*}
        \end{table}

        \begin{table*}[htb]
            \caption{ Partial widths of the M1 transitions for the $S$, $P$, and $D$ wave $B_c$ states compared with the other model predictions. }\label{M11}
            \begin{tabular}{cccccccccccccc}  \midrule[1.0pt]\midrule[0.5pt]
             ~~~Initial~~~  & ~~~Final~~~& \multicolumn{5}{c} {\underline{~~~~~~~~~~~~~~~~~~~~~~~~~~$E_{\gamma}$  (MeV)~~~~~~~~~~~~~~~~~~~~~~~~~~}} & \multicolumn{7}{c} {\underline{~~~~~~~~~~~~~~~~~~~~~~~~~~~~~~~~~~~~~~~~~~~~~$\Gamma_{\mathrm{M1}}$  (eV)~~~~~~~~~~~~~~~~~~~~~~~~~~~~~~~~~~~~~~~~~~~~~~~}}  \\
               ~~~state~~~ & ~~~state~~~  &~~~\cite{Eichten:1994gt}~~~&~~~\cite{Ebert:2002pp}~~~&~~~\text{GI\cite{Godfrey:2004ya}}~~~& ~~~\cite{Fulcher:1998ka}~~~&~~~Ours~~~
               &~~~\cite{Eichten:1994gt}~~~& ~~~\cite{Ebert:2002pp}~~~&~~~\text{GI \cite{Godfrey:2004ya}}~~~
               & ~~~\cite{Fulcher:1998ka}~~~ &~~~\text{MGI\cite{Li:2023wgq}}~~~ &~~ \cite{Martin-Gonzalez:2022qwd}~~&~~ Ours~~  						\\								
                 \midrule[0.5pt]	          %       EJ       EBERT       GOD     VVK      ME      EJ         EBERT        GOD        VVK     MGI      MG       ME
            $1 ^3S_{1}$	       &	$1 ^1S_{0}$   	&	72 	&	62      &	67	&	55	&	47	&	134.5	&	73  	&	80  	&	59   &  83.6  &   52  &	40.4	    \\
            $2 ^3S_{1}$	       &	$2 ^1S_{0}$   	&	43	&	46	    &	32	&	32	&	35	&	28.9	&	30  	&	10  	&	12   &  8.3   &   10  &	3.3	    \\
                               &	$1 ^1S_{0}$   	&	606	&	584	    &   588	&	599	&  604	&	123.4	&	141 	&	600 	&	122  &  559.3 &   650 & 562	    \\
            $2 ^1S_{0}$ 	   &	$1 ^3S_{1}$   	&	499	&	484	    &	498	&	520	&  528	&	93.3	&	160 	&	300 	&	139  &  320.6 &	  250 & 144     \\
            $1 ^3P_{2}$	       &	$1P_{1}$   	&		&	   	    &		&	 	&		&	     	&	    	&	    	&	     &        &	  & 0.13\\
                               &	$1P_{1}^{\prime}$   	&		&	   	    &		&	 	&		&	     	&	    	&	    	&	     &        &	  & 3.2 \\
                             
            $2 ^3P_{2}$        &	$1P_{1}$   	&		&	   	    &		&	 	&		&	     	&	    	&	    	&	     &        &	    & 15.7\\
                              &	$1P_{1}^{\prime}$   	&		&	   	    &		&	 	&		&	     	&	    	&	    	&	     &        &	    & 131.7\\
                               &	$2P_{1}$   	&		&	   	    &		&	 	&		&	     	&	    	&	    	&	     &        &	      & 0.10\\
                               &	$2P_{1}^{\prime}$   	&		&	   	    &		&	 	&		&	     	&	    	&	    	&	     &        &	    & 2.3\\
             $1D_{2}^{\prime}$	       &	$1^3D_{1}$   	&		&	   	    &		&	 	&		&	     	&	    	&	     	&	     &        &	  & 0.1\\
                                       &	$1^3D_{3}$   	&		&	   	    &		&	 	&		&	     	&	    	&	     	&	     &         &	& 0.022\\
            $1 ^3D_{3}$        &	$1D_{2}$   	&		&	   	    &		&	 	&		&	     	&	    	&	    	&	     &        &	  & 0.018\\
                             
            \midrule[0.5pt]\midrule[1.0pt]
            \end{tabular}
            \end{table*}

Following the preparation of the above parameters, we calculate the decay rates of the spin-nonflip $\pi\pi$ hadronic transitions and compare the numerical results with other Refs. \cite{Li:2023wgq,Godfrey:2004ya,Martin-Gonzalez:2022qwd} in Table \ref{E1-E1 transitions}, which includes our predictions for the processes of $B_{c}(2S) \to B_{c}(1S)\pi\pi$, $B_{c}(2P) \to B_{c}(1P)\pi\pi$ and $B_{c}(1D) \to B_{c}(1S)\pi\pi$. For the $B_{c}(2S) \to B_{c}(1S)\pi\pi$ process, it can be seen that the difference in the spin-nonflip $\pi\pi$ hadronic transition rates between spin triplets and spin singlets is relatively small, within about 10 keV. Our results are larger than the other theoretical values in Refs.~\cite{Li:2023wgq,Martin-Gonzalez:2022qwd}. Furthermore, it should be noted that the authors of Ref. \cite{Li:2023wgq} only calculated the hadron transition rates of the spin-nonflip $\pi^{+}\pi^{-}$, so the result is smaller than the results of ours and those in Refs. \cite{Godfrey:2004ya,Martin-Gonzalez:2022qwd}. Although we used the method in Ref. \cite{Godfrey:2004ya} to fit the transition rate of the process $B_{c}(2^{3}S_{1}) \to B_{c}(1^{3}S_{1})\pi\pi$ for $B_{c}$ states, there is still a difference of 4 keV compared to the results in Ref. \cite{Godfrey:2004ya}. And the transition rate of the process  $B_{c}(2^{1}S_{0}) \to B_{c}(1^{1}S_{0})\pi\pi$ is about 10 keV smaller than the results in Ref. \cite{Godfrey:2004ya}. In general, we find that the decay rates of the processes $B_{c}(2S) \to B_{c}(1S)\pi\pi$ are relatively high which are worth exploring experimentally.
%means that the possibility of discovery in the experiment is relatively high.

We can find a surprising conclusion from the processes of $B_{c}(2P)\to B_{c}(1P)\pi\pi$. Although all processes, except for $B_{c}(2^3P_{0}) \to B_{c}(1^3P_{0})\pi\pi$, $B_{c}(2P_{1}) \to B_{c}(1^3P_{2})\pi\pi$, $B_{c}(2^3P_{2}) \to B_{c}(1^3P_{0})\pi\pi$, $B_{c}(2^3P_{2}) \to B_{c}(1P_{1})\pi\pi$, and $B_{c}(2^3P_{2}) \to B_{c}(1^3P_{2})\pi\pi$, are suppressed, it can still be concluded from these significant decay rates that our numerical results are higher than those in Refs.~\cite{Godfrey:2004ya,Martin-Gonzalez:2022qwd}, which is caused by our adjustment of parameter $C_{2}$.
%Although it is difficult, it is undeniable that there is still a possibility that it can be detected experimentally.

Finally, it can be observed from the processes $B_{c}(1D) \to B_{c}(1S)\pi\pi$ we predicted above that our results are higher than those in Refs.~\cite{Godfrey:2004ya,Martin-Gonzalez:2022qwd,Li:2023wgq}. Furthermore, when comparing the processes of $B_{c}(1D_{2}) \to B_{c}(1^1S_{0})\pi\pi$ and $B_{c}(1D_{2}) \to B_{c}(1^3S_{1})\pi\pi$, we find that their numerical results are very similar, as demonstrated in Ref.~\cite{Godfrey:2004ya}.
Actually, Ref.~\cite{Godfrey:2004ya} reported higher values for the processes $B_{c}(1^3D_{1}) \to B_{c}(1^3S_{1})\pi\pi$ and $B_{c}(1^3D_{3}) \to B_{c}(1^3S_{1})\pi\pi$ than  those in Refs.~\cite{Li:2023wgq,Martin-Gonzalez:2022qwd}. In fact, only the transition rate of bottomonium has been considered to determine $C_{2}$ in the Refs.  \cite{Godfrey:2004ya,Martin-Gonzalez:2022qwd}, while we consider the fitting results of both charmomium and bottomonium as input, so the transition rates of $B_{c}(1D) \to B_{c}(1S)\pi\pi$ are higher than the results of these references. Thus, the transition rates of the processes $B_{c}(1D) \to B_{c}(1S)\pi\pi$ still need further experimental verification.

%The calculated results indicate that the measurement of the transition processes of $D$ wave states is very difficult.

\subsection{The analysis of $M$1 transitions}
\label{sec:AOM}
We have sorted out the predicted values of the $M$1 transitions for the $1S$, $2S$, $1P$, $2P$, and $1D$ wave of $B_{c}$ states and compared them with other Refs.~\cite{Li:2023wgq,Godfrey:2004ya,Eichten:1994gt,Ebert:2002pp,Martin-Gonzalez:2022qwd,Fulcher:1998ka} as shown in Table \ref{M11}. 
From the change in photon energy ($E_{\gamma}$), it is not difficult to see that our results are relatively similar to others', the difference being within tens of MeV, while it can also be seen that the mass spectrum of several theoretical models are within a reasonable range, since the photon energy depends on the mass spectrum.
In comparison, except for the processes of $B_{c}(1^3S_{1}) \to B_{c}(1^1S_{0})\gamma$ and $B_{c}(2^3S_{1}) \to B_{c}(2^1S_{0})\gamma$, 
our results are consistent with or larger than those of other references.
In fact, the E1 transition rates of the $B_{c}$ states are much larger than the $M$1 transition rates~\cite{Godfrey:2004ya,Li:2023wgq,Martin-Gonzalez:2022qwd}, but we can still draw some conclusions from the comparisons:

(i) Compared with $P$ and $D$ waves, the $M$1 transition rates of the $S$ waves are generally larger, and the $S$ wave states $B_{c}(2^3S_{1})$ and  $B_{c}(1^3S_{1})$ are worth discussing, which are conducive to the determination of their values in experiment.

(ii) The $M$1 transition rate of the processes of $B_{c}(2^3S_{1}) \to B_{c}(1^1S_{0})\gamma$ is calculated in this work is similar to those compared with Refs.~\cite{Godfrey:2004ya,Li:2023wgq}, but about 400 eV higher than those in Refs.~\cite{Eichten:1994gt,Ebert:2002pp,Fulcher:1998ka}.  

(iii) %The  dipion transition is further suppressed by the contribution of colour magnetic interactions. 
%\sout{This is worth considering whether such a contribution relationship also exists in the $B_{c}$ states.}
Based on the calculated data, we have provided the prediction value of product branching fraction ${\cal B} (B_{c}(2^3S_{1}) \to \gamma B_{c}(2^1S_{0}))\times{\cal B}(B_{c}(2^1S_{0})\to \pi\pi B_{c}(1^1S_{0}))$ as $4.31\times 10^{-5}$.

In general, the $M$1 transition rate of the process of $B_{c}(2^3S_{1}) \to B_{c}(1^1S_{0})\gamma$ in the $S$ wave states is comparatively the highest at about 500 eV, and if the $M$1 transition is to be used to determine $B_{c}(2^3S_{1})$, this process is undoubtedly the best choice. However, according to our prediction, if there is a contribution of $M$1 radiative transition to the dipion transition in the $B_{c}$ states, the process of $B_{c}(2^3S_{1}) \to B_{c}(2^1S_{0})\gamma$ can also be explored to determine $B_{c}(2^3S_{1})$, although the $M$1 radiative transition rate of the process of $B_{c}(2^3S_{1}) \to B_{c}(2^1S_{0})\gamma$ is very small.

\section{Summary}
\label{sec:summary}

So far, the $B_{c}$-meson family remains to be further explored.
%, and we still have very little research on its properties and mechanism.
In this paper, we mainly studied the  decay rates of spin-nonflip $\pi \pi$ and  $M$1 radiative transitions of $B_{c}$ states based on the Cornell potential model with a screening potential effect.
%In particular, for the spin-nonflip $\pi \pi$ transition of the $B_{c}$ states, 
%an experimental value of 57 keV has been included in the PDG \cite{ATLAS:2014lga}. 
%{\color{orange}Although different models provide different predicted values, our results are also within a reasonable range.}

We have adopted the Kuang-Yan approach, the QCDME method to calculate the process of the spin-nonflip $\pi \pi$ transitions,
and the QCS model to calculate the spectrum of the intermediate hybrid mesons. Our results about the $S$ wave states are basically consistent with those in Refs. \cite{Godfrey:2004ya,Martin-Gonzalez:2022qwd}.
However, for $P$ wave states, the process of the $B_{c}(2^3P_{0}) \to B_{c}(1^3P_{0})\pi\pi$ transition has a decay rate of 104 keV, which is higher than those in Refs.~\cite{Godfrey:2004ya,Li:2023wgq,Martin-Gonzalez:2022qwd}. As for $D$ wave states, our results are higher than those in other literature \cite{Godfrey:2004ya,Li:2023wgq,Martin-Gonzalez:2022qwd} as well, due to the consideration of both charmonium and bottomonium. Although the $M$1 radiative transition rates are very small, we can give some useful information for $B_{c}$ states. Taking inspiration from Ref.~\cite{BESIII:2024mdm}, we have given the prediction value of the product branching fraction ${\cal B} (B_{c}(2^3S_{1}) \to \gamma B_{c}(2^1S_{0}))\times{\cal B}(B_{c}(2^1S_{0})\to \pi\pi B_{c}(1^1S_{0}))$ as $4.31\times 10^{-5}$, although this prediction still needs further data and experimental verification. The determination of the $B_{c}(2^3S_{1})$ state may be achieved through the processes of $B_{c}(2^3S_{1}) \to B_{c}(1^1S_{0})\gamma$ and $B_{c}(2^3S_{1}) \to B_{c}(2^1S_{0})\gamma$.

All in all, we expect that our numerical results will provide some reference for the study of the properties of $B_{c}$ states and make some contributions to further studies.

\acknowledgments

\label{sec:ack}
We are very grateful for professor Zhan-Wei Liu's useful discussion. This work is supported by the National Natural Science Foundation of China under Grants No. 11965016 and No. 12247101; and by the Natural Science Foundation of Qinghai Province under Grant No. 2022-ZJ-939Q.

\appendix
\section{THEORETICAL MODELS OF MASS SPECTRUM}
\label{sec:TmoMS}
This appendix is a brief introduction to the nonrelativistic Cornell potential model with a screening potential effect. In the nonrelativistic case, the Hamiltonian of the model is 
\begin{align}\label{2.1}
H=H_{0}+V,
\end{align}
and $H_0$ denotes
\begin{align}
%H_{0} \rightarrow \sum_{i=1}^{2}\left(m_{i}+\frac{p^{2}}{2 m_{i}}\right),
H_{0} = \sum_{i=1}^{2}\left(m_{i}+\frac{p^{2}}{2 m_{i}}\right),
\end{align}
where ${m_{i}}(i=1,2)$ are the masses of $\bar{b}$ and $c$ quarks, respectively. And for the Cornell potential  \cite{Eichten:1978tg}, 
\begin{align}
G(r)=-\frac{4 \alpha_{s}}{3 r},  \\   
s(r)=br+c, 
\end{align}
where $G(r)$ and $s(r)$ are the Coulomb and linear potentials, respectively, and the parameter $c$ denotes the scaling parameter \cite{Eichten:1979ms}. When considering the screening effect \cite{Pang:2017dlw}, the linear potential can be changed to
\begin{align}
s(r)^{\prime}=\frac{b\left(1-e^{-\mu r}\right)}{\mu}+c,
\end{align}
where $\mu$ is a screening parameter.

For the form of the spin-dependent term, reference was made to the Godfrey-Isgur (GI) model \cite{Godfrey:1985xj,Lang:1982tj,Michael:1992nj,Bali:2000gf,Kawanai:2011xb}. And it makes corresponding corrections to the spin correlation term of the linear potential after incorporating the screening effect. Thus, we have
\begin{align}
V=H^{conf}+H^{cont}+H^{so}+H^{ten},\label{2.6}
\end{align}
in which $H^{conf}=G(r)+s(r)^{\prime}$ 
contains the Coulomb-like and screening potential interaction. The color contact interaction can be written as
\begin{align}
H^{cont}=\frac{32 \pi \alpha_{s}}{9 m_{1} m_{2}}\left(\frac{\sigma}{\pi^{\frac{1}{2}}}\right)^{3} e^{-\sigma^{2} r^{2}} \vec{S}_{1} \cdot\vec{S}_{2}.
\end{align}
The third term
\begin{align}
H^{so}=H^{so(cm)}+H^{so(tp)}
\end{align}
is the spin-orbit interaction, where
\begin{align}
H^{so(cm)}=\frac{4 \alpha_{s}}{3} \frac{1}{r^{3}}\left(\frac{1}{m_{1}}+\frac{1}{m_{2}}\right)^{2} \vec{L} \cdot \vec{S}_{1(2)}
\end{align}
and
\begin{align}
H^{so(tp)}&=-\frac{1}{2 r} \frac{\partial H^{c o n f}}{\partial r}\left(\frac{{\vec{S}}_{1}}{m_{1}^{2}}+\frac{{\vec{S}}_{2}}{m_{2}^{2}}\right) \cdot {\vec{L}}\nonumber\\
&=-\frac{1}{2 r}\left(\frac{4 \alpha_{s}}{3} \frac{1}{r^{2}}+b e^{-\mu r}\right)\left(\frac{1}{m_{1}^{2}}+\frac{1}{m_{2}^{2}}\right) \vec{L} \cdot \vec{S}_{1(2)}
\end{align}
is the Thomas precession term with the screening effect. Additionally, we define 
\begin{align}
H^{ten}=\frac{4}{3} \frac{\alpha_{s}}{m_{1} m_{2}} \frac{1}{r^{3}}\boldsymbol{T},
\end{align}
which depicts the color tensor interaction, and
\begin{align}
&\boldsymbol{T}=\frac{ 3\vec{S}_{1} \cdot \vec{r} \vec{S}_{2} \cdot \vec{r}}{{\vec{r}}^{2}}-\vec{S}_{1} \cdot \vec{S}_{2},\\
&\langle \boldsymbol{T}\rangle= \begin{cases}-\frac{L}{6(2 L+3)} & J=L+1 \\ \frac{1}{6} & J=L \\ -\frac{(L+1)}{6(2 L-1)} & J=L-1\end{cases}
\end{align}
where $\boldsymbol{T}$ is the tensor operator,  $\vec{S}_{1}$ and $\vec{S}_{2}$ are the spins of the quark and antiquark contained by the meson, and $\vec{L}$ is the orbital angular momentum \cite{Kawanai:2011fh}.

The eigenvalues and eigenvectors of the mass spectrum of $B_{c}$ states are calculated by using the simple harmonic oscillator (SHO) base expanding method. In configuration and momentum space, SHO wave functions have explicit form, respectively,

\begin{align}
&\Psi_{n L M_{L}}(\boldsymbol{r})=R_{n L}(r, \beta) Y_{L M_{L}}\left(\Omega_{r}\right), \\
&\Psi_{n L M_{L}}(\boldsymbol{p})=R_{n L}(p, \beta) Y_{L M_{L}}\left(\Omega_{p}\right),
\end{align}
where
\begin{align}
&R_{n L}(r, \beta)=\beta^{\frac{3}{2}} \sqrt{\frac{2 n !}{\Gamma\left(n+L+\frac{3}{2}\right)}}(\beta r)^{L} e^{\frac{-r^{2} \beta^{2}}{2}} L_{n}^{L+\frac{1}{2}}\left(\beta^{2} r^{2}\right), \\
&R_{n L}(p, \beta)=\frac{(-1)^{n}(-i)^{L}}{\beta^{\frac{3}{2}}} e^{-\frac{p^{2}}{2 \beta^{2}}} \sqrt{\frac{2 n !}{\Gamma\left(n+L+\frac{3}{2}\right)}}\left(\frac{p}{\beta}\right)^{L} L_{n}^{L+\frac{1}{2}}\left(\frac{p^{2}}{\beta^{2}}\right)\blue{,}
\end{align}
where $Y_{L M_{L}}\left(\Omega_{r}\right)$ is a spherical harmonic function, $R_{n L}$ ($n=0,1,2,3,\ldots$) is a radial wave function, and $L_{n}^{L+\frac{1}{2}}(x)$ denotes a Laguerre polynomial.

The introduction of the truncation parameter $r_{c}$ can reasonably consider the correction of the mass spectrum and wave function of $B_{c}$ states by the spin-orbit and tensor terms, and can overcome the singular behavior of $1/r^{3}$ in these two terms. In addition, this method has successfully processed the mass spectra of $b\bar{b}$, $c\bar{c}$, and $c\bar{b}$ \cite{Deng:2016stx,Deng:2016ktl,Li:2019tbn}. In a small range $r\in(0,r_{c})$, we set $1/r^{3}=1/r_{c}^{3}$.

\section{THE CALCULATION PROCESS OF SPIN-NONFLIP \texorpdfstring{$\pi\pi$}{} TRANSITION}
\label{sec:st}
In this appendix, we provide specific calculation processes for the spin-nonflip $\pi\pi$ transition and explanations. The transition rate is given by \cite{Wang:2018rjg}

\begin{equation}
\begin{split}
\Gamma\left(\Phi_{i} \to
\Phi_{f} + \pi\pi\right) &=
\delta_{l_{i}l_{f}}\delta_{J_{I}J_{F}} (G|C_{1}|^{2}-\frac{2}{3}H|C_{2}|^{2}
)\left|\mathcal{A}_{1}\right|^{2} \\
&
+(2l_{i}+1)(2l_{f}+1)(2J_{f}+1)\\
&
\times \sum_{k} (2k+1) (1+(-1)^{k})
\left\lbrace\begin{matrix} s & l_{f} & J_{f} \\ k & J_{i} & l_{i}
\end{matrix}\right\rbrace^{2} \\
&
\times H |C_{2}|^{2}\left|\mathcal{A}_{2} \right|^{2},
\label{eq:gamapipi}
\end{split}
\end{equation}
with
\label{sec:Ht}
\begin{equation}
    \begin{split}
   \mathcal{A}_{1}=& \sum_{l}(2l+1) \left(\begin{matrix} l_{i} & 1 & l \\ 0 & 0 & 0
    \end{matrix}\right) \left(\begin{matrix} l & 1 & l_{i} \\ 0 & 0 & 0
    \end{matrix}\right) f_{if}^{l11},\\
    \mathcal{A}_{2}=&\sum_{l} (2l+1) \left(\begin{matrix} l_{f} & 1 & l \\ 0 & 0 & 0
    \end{matrix}\right) \left(\begin{matrix} l & 1 & l_{i} \\ 0 & 0 & 0
    \end{matrix}\right) \left\lbrace\begin{matrix} l_{i} & l & 1 \\ 1 & k & l_{f}
    \end{matrix}\right\rbrace f_{if}^{l11},
    \label{eq:A}
\end{split}
    \end{equation}
where
\begin{equation}
\begin{split}
f_{if}^{lP_{i}P_{f}} &= \sum_{k} \frac{1}{M_{i}-M_{kl}} \left[\int dr\,
r^{2+P_{f}} R_{f}(r)R_{kl}(r)\right] \\
&
\times \left[\int dr' r'^{2+P_{i}} R_{kl}(r')
R_{i}(r')\right].
\label{eq:fifl}
\end{split}
\end{equation}
$R_{kl}(r)$ is the radial wave function of the intermediate quark-gluon states, whereas $R_{i}(r)$ and $R_{f}(r)$ are the radial wave functions of the initial and final states, respectively. The mass of the decaying meson is $M_{i}$, whereas the ones corresponding to the hybrid states are $M_{kl}$. The quantities $G$ and $H$ are phase-space integrals
\begin{equation}
\begin{split}
G=&\frac{3}{4}\frac{M_{f}}{M_{i}}\pi^{3}\int
dM_{\pi\pi}^{2}\,K\,\left(1-\frac{4m_{\pi}^{2}}{M_{\pi\pi}^{2}}\right)^{1/2}(M_{
\pi\pi}^{2}-2m_{\pi}^{2})^{2}, \\
H=&\frac{1}{20}\frac{M_{f}}{M_{i}}\pi^{3}\int
dM_{\pi\pi}^{2}\,K\,\left(1-\frac{4m_{\pi}^{2}}{M_{\pi\pi}^{2}}\right)^{1/2}
\times \\
&
\times\left[(M_{\pi\pi}^{2}-4m_{\pi}^{2})^{2}\left(1+\frac{2}{3}\frac{K^{2}}{M_{
\pi\pi}^{2}}\right)\right. \\
&
\left.\quad\,\, +\frac{8K^{4}}{15M_{\pi\pi}^{4}}(M_{\pi\pi}^{4}+2m_{\pi}^{2}
M_{\pi\pi}^{2}+6m_{\pi}^{4})\right],
\end{split}
\end{equation}
with the momentum $K$ given by
\begin{equation}
K = \frac{\sqrt{\left[(M_{i}+M_{f})^{2}-M_{\pi\pi}^{2}\right]
\left[(M_{i}-M_{f})^{2}-M_{ \pi\pi}^{2}\right]}}{2M_{i}},
\end{equation}
in which $m_{\pi}$ is the mass of pion and the invariant mass lying the interval $4m_{\pi}^{2}\leq M_{\pi\pi}^{2}\leq (M_{i}-M_{f})^{2}$.

\section{THE INTRODUCTION AND EFFECTIVE POTENTIAL FOR HYBRID MESONS}
\label{sec:ep}

In this appendix,{\color{black} we briefly introduce the QCS model and provide the effective potential of the hybrid states. 

The meson is composed of quark and antiquark that are connected by an appropriate color electric flux tube (string). The string can carry energy momentum only in the region between the quark and the antiquark. Even the string and the quark-antiquark pair as a whole can rotate or vibrate. When not considering the vibration of string, the dynamics of the string, the quark, and the antiquark can be described using the Schr$\ddot{\mathrm{o}}$dinger equation with a constrained potential.}
When the string vibrates, the gluon excitation effect is taken into account, 
which describes a new state of gluons and quark-antiquarks, known as a hybrid state.

The effective potential for hybrid mesons can be expressed as \cite{Buchmuller:1979gy}
\begin{equation}
V_{\rm hyb}(r)=V_{\rm G}(r) + V_{\rm S}(r) +
\left[V_{n}(r) - \sigma(r)r\right],
\label{eq:pothyb}
\end{equation}
where $V_{G}(r)$ is one-gluon exchange potential and $V_{S} (r)$ is a color confining potential. 
It is obvious that when ${\it n} = 0$ in $V_{n}$, the effective potential becomes a descriptive potential of a quark-antiquark pair. Here, we talk about the hybrid meson and take ${\it n}=1$. And then the effective vibrational potential can be given by \cite{Giles:1977mp}
\begin{equation}
V_{n}(r) = \sigma(r)r \left\lbrace 1 - \frac{2n\pi}{2n\pi+ \sigma(r)\left[(r-2d)^{2}+4d^{2}\right]} \right\rbrace^{-1/2},
\end{equation}
where 
\begin{equation}
    \sigma(r) =\frac{b(1-e^{-\mu r})}{\mu r},
\end{equation}
where the vibrational potential energy can be estimated using the Bohr-Sommerfeld quantization and assuming the quark mass to be very heavy so that the ends of the string are fixed~\cite{Giles:1977mp}. To relax the last assumption one can define a parameter $d$ given by 
\begin{equation}
  d(m_{b},m_{c},r,\sigma,n) = \frac{\sigma(r)r^{2}\alpha_{n}}{4(m_{b}+m_{c}+\sigma(r)r\alpha_{n})},
\end{equation}
in which $d$ is the correction of the finite heavy quark mass. $\alpha_{n}$ relates to the shape of the vibrating string~\cite{Giles:1977mp} and can take the values $1\leq\alpha_{n}^{2}\leq2$. We take $\alpha_{1}=\sqrt{1.5}$.

Because of the screening potential effect, the effective string tension $\sigma(r)r$ is a function of the distance $r$ between $c$ and $\bar{b}$.
For theoretical self-consistency, the form of $V_{S} (r)$ is taken from the Cornell potential model with the screening potential.
The specific potential $V_{G} (r)$ and $V_{S} (r)$ are given by
\begin{equation}
\begin{split}
V_{\rm G}(r) &= -\frac{4\alpha_{s}}{3r}, \\
V_{\rm S}(r) &= \sigma (r)r + c. 
\end{split}
\end{equation}   

The threshold of the hybrid potential is defined as
\begin{equation}
 V_{\rm hyb}(r)\xrightarrow{r\rightarrow \infty} \frac{b}{\mu}+c.
\end{equation}

\bibliographystyle{apsrev4-1}
\bibliography{ref}

\end{document}